# 5G Core Fault Detection and Root Cause Analysis using Machine Learning and Generative AI

Joseph H. R. Isaac,[a)] Harish Saradagam,[b)] and Nallamothu Pardhasaradhi[c)]

*Global AI Accelerators (GAIA)*
*Ericsson India.*
[a)]Corresponding author: joseph.isaac@ericsson.com
[b)]harish.s@ericsson.com
[c)]nallamothu.pardhasaradhi@ericsson.com

**Abstract:** With the advent of 5G networks and technologies, ensuring the integrity and performance of packet core traffic is paramount. During network analysis, test files such as Packet Capture (PCAP) files and log files will contain error if present in the system that must be resolved for better overall network performance such as connectivity strength and handover quality. Current methods require numerous person-hours to sort out testing results and find the faults. This paper presents a novel AI/ML-driven Fault Analysis (FA) Engine designed to classify successful and faulty frames in PCAP files, specifically within the 5G packet core. The FA engine analyses network traffic using natural language processing techniques to identify anomalies and inefficiencies, significantly reducing the effort time required and increasing efficiency. The FA Engine also suggests steps to fix the issue using Generative AI via a Large Language Model (LLM) trained on several 5G packet core documents. The engine explains the details of the error from the domain perspective using documents such as the 3GPP standards and user documents regarding the internal conditions of the tests. Test results on the ML models show high classification accuracy on the test dataset when trained with 80-20 splits for the successful and failed PCAP files. Future scopes include extending the AI engine to incorporate 4G network traffic and other forms of network data, such as log text files and multimodal systems.

## INTRODUCTION

As the telecommunications industry continues to advance, the integration of cutting-edge technologies and the intricate interactions among network components present considerable challenges for network troubleshooting. Software testing is a critical phase in telecommunications that ensures the quality, functionality, and reliability of network products [1]. It involves executing network systems to identify defects, bugs, errors, issues, or unmet requirements, collectively referred to as faults. The primary goal is to detect these faults early in the development process, thereby minimizing the cost and effort associated with fixing them in later stages [2]. By analysing the faults, the root cause must be found and fixed at the earliest to improve the system's performance and throughput. In this paper, the focus will be on PCAP files and root cause analysis using these files.

A PCAP file is a data file that contains captured network traffic, which is collected and stored by network monitoring software such as Wireshark [3], tcpdump, or similar tools. The file records the details of network packets transmitted and received over a network, including headers and payload data. Each packet captured is timestamped. and includes information like source and destination IP addresses, port numbers, protocol types, and packet contents. This information is essential for network diagnostics, troubleshooting, performance analysis, and security monitoring, as it allows network administrators and analysts to review and analyse network communication at a granular level, identifying issues such as latency, packet loss, or malicious activity. The massive volume of data and messages exchanged within these networks, coupled with the urgent need for accurate and timely fault detection, calls for modern approaches that surpass the limitations of traditional methods.

One of the challenges of manual testing is that a considerable amount of person-hours is required to comprehensively evaluate network PCAP files. This delay arises from various factors, primarily the increasing complexity of modern telecommunications systems. Human testers must execute test cases, document outcomes, and verify results. These are tasks that are inherently time-consuming and susceptible to human error. This increases the risk of overlooking critical defects, potentially compromising software quality. As network applications grow more intricate, they demand extensive test cases covering a wide range of functionalities, configurations, scenarios, and edge cases [4]. Designing, executing, and analysing each test case requires meticulous attention, specialized knowledge, and labour-intensive in nature. By reducing manual intervention, software development teams can enhance testing efficiency while ensuring software reliability and performance.

## Proposed Solution

In order to improve the performance and time efficiency of network based software testing, we propose the 5G Core Fault Analysis Engine, which can automatically find faults in PCAP files and point them out to the user, along with the steps for understanding the fault with possible resolutions. The engine is trained on a diverse dataset of labelled PCAP files, encompassing various scenarios of normal and anomalous traffic. During inference, the FA engine will analyze any PCAP file and point out the faulty frames of the PCAP file. The FA engine also utilizes Generative AI via a Large Language Model (LLM). LLMs represent a significant advancement in artificial intelligence, particularly in natural language processing (NLP). These models are designed to understand, generate, and interact with human language to mimic human-like understanding and response generation. They are deep learning models trained on vast amounts of text data to perform various language-related tasks. These models have billions of parameters, allowing them to capture intricate patterns, contextual information, and semantic nuances in the text.

To enhance the effectiveness of the troubleshooting step, the LLM is provided with a retrieval-augmented generation (RAG) which is an external retrieval mechanism. This approach allows the model to access relevant information from external knowledge bases or documents, ensuring more accurate and contextually enriched responses. In our proposed solution, the RAGs are created using 5G packet core-related documents such as the 3GPP standards [5] and the Packet Core Controller (PCC) [6] to provide domain-based knowledge on the fault and the required steps to fix the fault. The Packet Core Controller (PCC) is a cutting-edge, cloud-native control plane signalling processing function within the 5G Core. It is designed to handle access, session, mobility, and gateway control functions, supporting the latest 5G use cases. Compliant with the 3GPP Release 15 specifications, this controller implements essential control plane network functions, including the MME, SGW-C, PGW-C, AMF, and SMF. It has advanced features such as reduced signalling, adaptive paging algorithms, and the ability to maintain service continuity for subscribers during network disturbances. Additionally, it offers network assurance with software probes, making it a comprehensive solution for Mobile Broadband (MBB), Massive IoT, VoLTE, and 5G NSA/SA deployments.

Overall the core contributions of our proposed solution include:
- An accurate fault detection system that will find the faulty frames based on both data driven and machine learning based approaches.
- A robust pre-processing method to extract the data from the PCAP files for analysis.
- A system that provides troubleshooting steps for the detected fault to circumvent the error based on prior knowledge as well as domain knowledge.
- Can be easily scaled to incorporate various test scenarios and call flows.

## RELATED WORKS

Root cause analysis is a well-established research area with extensive literature addressing various applications [7, 8, 9, 10]. However, its application to 5G telecommunications networks remains relatively underexplored due to the unique complexities and dynamic nature of these networks. Fault analysis in 5G telecommunications has attracted considerable research interest, with numerous systems leveraging advanced machine learning techniques to address the challenges inherent in managing such highly interconnected systems [11, 12, 13].

One notable contribution in this space is the work by Mfula et al. [14], who proposed an adaptive root cause analysis (ARCA) framework based on Bayesian network theory. This method automates the process of determining root causes by building probabilistic models that capture dependencies among network components, enabling automated fault detection and diagnosis.

Further advancements are exemplified by Munoz et al. [15], who developed a temporal analysis-based methodology for identifying root causes in self-organizing 5G networks. Their approach employs time-series data analysis to detect abnormal patterns and link them to potential root causes, enhancing fault detection in complex, dynamically evolving network environments.

Zhao et al. [16] introduced a fault diagnosis method that integrates knowledge-based systems with data fusion techniques. Their model synthesizes diverse data sources, including performance metrics, alarms, and logs, to improve fault diagnosis accuracy. Despite its promise, this approach requires significant computational resources and extensive model training, posing challenges for real-time deployment.

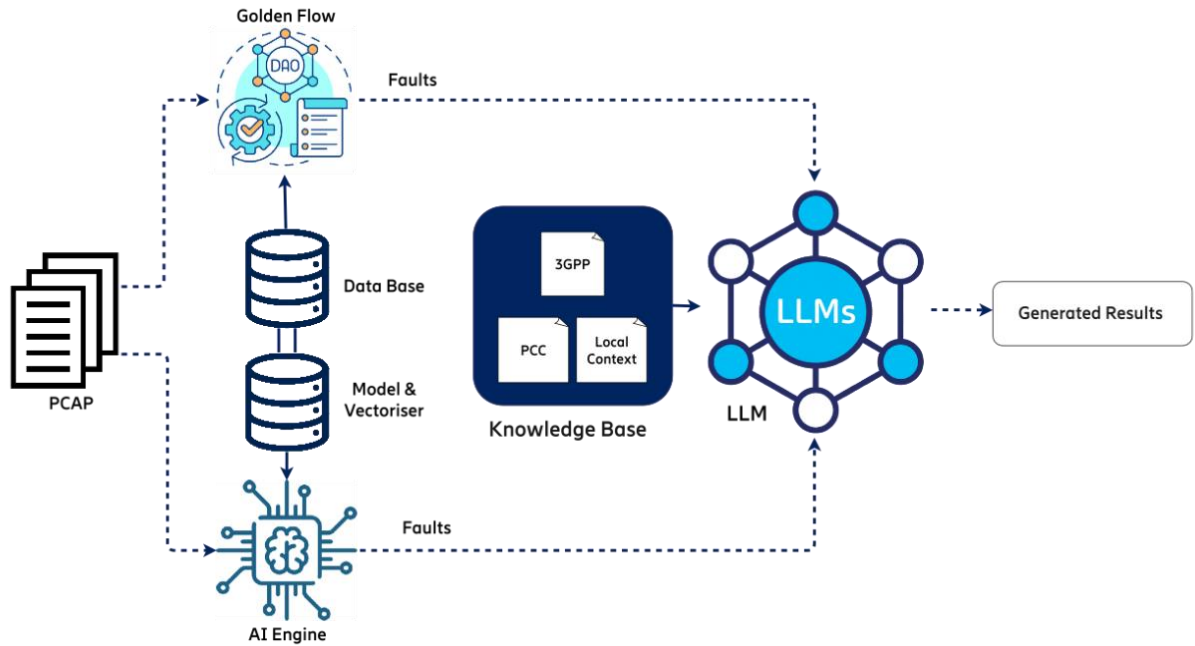

**FIGURE 1.** Architecture diagram for the 5G Fault Analysis Engine. The process begins with the PCAP file as the input. This file will first be fed to the AI engine and the Golden Flow model, which then processes the file and determines the faults present in the PCAP file. These errors and faults are then sent to the LLM, which is trained with the PCC, local contexts such as test documents and frequently performed test methods, and the 3GPP standards.

Another innovative method was introduced by Zhang et al. [17], who employed Word2Vec [18] to extract semantic features from 5G alarm information. They combined this feature set with the CatBoost algorithm to detect hidden patterns and predict solutions, demonstrating the potential of natural language processing techniques in fault diagnosis.

Despite these advancements, practical implementation challenges persist, particularly in large-scale testing environments. For example, during nightly test sessions involving multiple simultaneous test executions, testers must manually review extensive test logs and PCAP files the following day. This manual analysis is time-consuming, requiring a significant workforce and considerable effort [19]. Ideally, the testers require a system to automatically find the problem in the PCAP file when they open to analyse the file.

PCAP data is useful for diagnosing network problems and conducting in-depth analysis of network systems and call flow testing. Several methods were proposed aimed at detecting faults within network traffic captured through PCAP files [20, 21, 22]. Much of the existing work has centred on recognizing unusual traffic patterns that could signal network malfunctions or potential security breaches. These works relied heavily on manual inspections and rule-based detection systems. Although they are effective in locating errors, They can be labour-intensive and testers can struggle to keep up with the growing volume of network data. A related work using PCAPs is done by Tulczyjew et al. [23], who made a system called LLMcap, a self-supervised large language model (LLM) approach for detecting network failures in PCAP data. They used DistilBERT-based masked language modelling (MLM) and LLMcap identifies anomalies in network data without the need for labelled datasets. However, this method does not explain the anomaly; instead, it detects the anomaly alone. We need to detect and resolve the faults found in the PCAP files using domain knowledge.

## METHODOLOGY

The architecture diagram of the whole process is shown in Figure 1. The process begins with the PCAP file as the input. The data from the PCAP file is extracted using Wireshark libraries that converts the files to a readable json file that can be used by the system. This data will first be fed to the AI engine and the Golden Flow module, which then

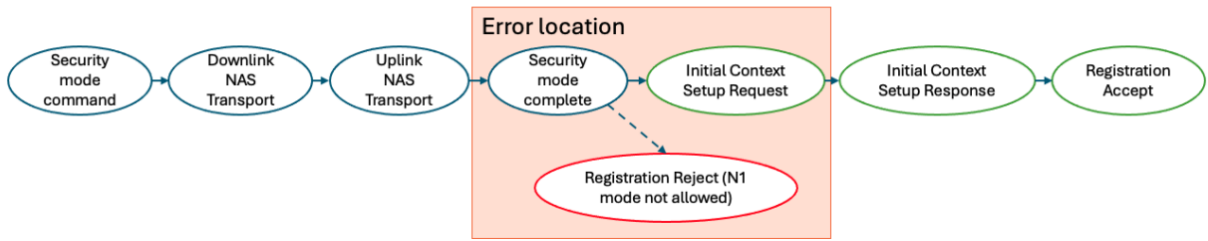

(a) Sample subset of PCAP file with an expected good test result

(b) Sample subset of PCAP file with a negative test result.

(c) Graph representation of the scenario.

**FIGURE 2.** Sample scenario representation of the Golden Flow model. From Figure 2c it is clear that there is a deviation where the system expected an "Initial Context Setup Request" but a registration reject message came instead. This location will be marked by the Golden flow model as the error location.

processes the file and determines the faults present in the PCAP file. These fault messages are then sent to an LLM, which generates the details of the fault, as well as steps to resolve the fault.

## Golden Flow

The golden flow model is a knowledge graph-based analysis method [24] that creates a graph based on the successful PCAP test files. The individual frames of the PCAP are nodes of the graph, and each frame in the PCAP file will be connected to its adjacent next frame with a directed edge. During inference, the frames of the PCAP file will be used to traverse the graph to find out if the file given is successful by reaching the end of the graph. The advantage of this method is that not only will the model find out if there is an anomaly from the learned data, but it can also suggest the correction. This suggestion can help the developer to understand the problem better.

For example, a few frames of a sample call flow PCAP file is shown in Figure 2. In Figure 2a and Figure 2b, a few frames for a simple registration call flow in 5G core is shown. The first shows how a call flow is supposed to be, while the second set has an error which resulted in a registration reject. Figure 2c is the graph created using the frames of successfully executed test run PCAP files. As seen in the graph, the flow is supposed to go to "InitialContextSetupRequest", but the flow diverges to a different node name (Registration Reject in this case). As this node is not seen by the graph, it will be marked as an anomaly and reported to the user along with the expected passages (there can be more than one legal move from a node in a well-trained graph model).

## AI Engine

The AI engine is described in Figure 3 and is an NLP model that reads and classifies the individual frames stored in PCAP files, determining whether they are faulty or correct. The main difference between the Golden Flow model and the AI Engine is that the former one is data-driven while the latter is a supervised machine learning model. The messages from the PCAP undergo preprocessing to prepare the data for analysis. This step involves cleaning the text by removing noise (e.g., headers, metadata, non-alphanumeric characters), tokenizing the text into individual words or phrases, and normalizing the text by converting it to lowercase. After preprocessing, the text is vectorized using the Bag-of-words method. These vectors capture the messages' semantic meaning and contextual information, enabling the model to analyse and compare them effectively. With the vectorized representations, an ML model is trained to classify the messages. The model used in this paper is the Support Vector Machine [25, 26], which has shown promising results in Table 2 in comparison to other models which are also shown in the results. To train the model, the user needs to provide two sets of PCAP files. The first set of PCAP files are known to be correct and from a successful test run. The second set of PCAP files are from failed test runs. The specific frames which are the cause of the error need not be labelled and will be found automatically by the system. The sets of PCAP files will be used to create a training set in which all the frames from the PCAPs in successful tests are considered correct frames as those frames did not result in any fault. To get the negative frames, any frame present in the PCAP file of failed tests that is not present in the successful PCAP files is considered as a negative or faulty message frame.

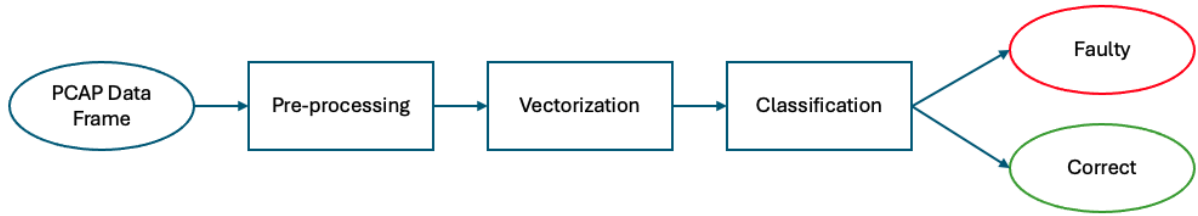

**FIGURE 3.** Inference process of the AI engine. The frame of the PCAP file will be pre-processed and then sent for vectorization. The vectorised data will then be used as input for the classification model which is the Support Vector Machine. The model will then determine whether the frame is faulty or not.

## LLM-based Troubleshooting Agent

After the Golden Flow and the AI Engine perform the inference and decide if the frame is faulty, the reason for the fault and corrective measures for the system will be needed. Hence an LLM is used to generate troubleshooting steps for identified faults in network systems. This paper uses MistralAI [27] as the LLM model and utilise semantic RAGs [28] to understand the problem in the frame and provide the root cause with the troubleshooting steps. The RAGs are created using a corpus of 5G testing schemes and error codes stored in a cloud platform.

In order to create the prompt for the LLM, the context will be automatically provided based on the frames of the PCAP file. For example, if the Golden Flow and AI engine name a frame faulty which has the message: "Registration Reject, Congestion", then a prompt will be created that will mention the content of this frame as well as the contents of the previous frame for extra context. The model will then use the context gained from the RAGs to find the relevant information and then give the suggested steps from the document as well as references to any related documents. The PCC specifications and cause codes are outlined in detailed documents from which the RAG is created. The software testers also prepare a document that contains frequently encountered problems and the steps taken by the testers to solve the error. The RAG will then be used to show any similar issue during inference. The LLM will use these RAGs to understand the problem context better and provide meaningful resolutions without hallucinating any answers.

## RESULTS AND DISCUSSION

To train the AI Engine and the Golden Flow model, 198 PCAP files were used, where 58 files were PCAPs of successful 5G core tests and 140 failed tests. The tests involve user mobile registration, PDU session establishment, and user deregistration. These PCAP files were created in a simulated lab environment. These files are then pre-processed as mentioned in Section and split into train and test sets with an 80/20 split. The Golden Flow detected every test file, resulting in 100% detection accuracy. The results of the AI Engine are shown in Table 1 and Table 2. In Table 2, the first column denotes the type of grouping performed before classification. For example, "NAS-5GC" means that a separate ML model was trained for this type of frame, which corresponded to the negative and positive samples. From the table, it is clear that combining the ML model learning from all protocols is not advisable as it impacts the recall of the system. To improve the system's performance, the frames are first categorized by the protocol type and then split into the training and the test sets. With this kind of categorization, the system showed improvement in the individual protocol types. Table 1 shows the comparison of various models tested for the final classifier that uses the vectorised values from the message. All experiments were performed with all protocol types put together. Based on the comparison, the support vector machine shown better results than the rest.

**TABLE 1.** To Results from the AI Engine model comparison. From the table it is clear that SVMs performed better than the rest of the models.

| Model | Accuracy | F1-Score |
|---|---|---|
| Support Vector Machine | 86% | 84% |
| Neural Network | 81% | 80% |
| Random Forest | 83% | 81% |
| Bayes Classifier | 79% | 76% |

**TABLE 2.** Results from the AI Engine test experiments. The first column denotes the protocol in which the frames were categorized before testing with the model. The support vector machine was used in all these configurations.

| Type | Accuracy | Precision(-ve) | Recall(-ve) | F1-Score |
|---|---|---|---|---|
| All protocols together | 86% | 90% | 79% | 84% |
| HTTP2 | 99% | 80% | 80% | 89% |
| HTTP2/JSON | 99% | 100% | 80% | 89% |
| HTTP2/JSON/NAS-5GC | 100% | 100% | 100% | 100% |
| NAS-5GC | 100% | 100% | 100% | 100% |

To test the LLM, a file which contains an error in the Registration call flow (one of the primary call flows in 5G core) was introduced to the model. The Golden Flow and AI Engine collectively provided the following results:
1) Suspicious Message: Registration reject (Congestion) at frame:23
2) Expected messages: InitialContextSetupRequest to InitialContextSetupResponse
3) Possible causes:
   a) 20 HEADERS[1]: 504 Gateway Time-out, DATA[1], JSON (application/problem+json)
   b) 22 HEADERS[3]: 504 Gateway Time-out, DATA[3], JSON (application/problem+json)

The LLM also provided a detailed procedure to fix the congestion error in the file. It also referred to documents related to the error so that users could understand the problem better.

## CONCLUSION AND FUTURE WORKS

In this paper, a novel 5G Core Fault Analysis Engine was proposed, which analyses frames in a PCAP file to detect the faults present using two modules, namely the Golden Flow and the AI engine. The two modules will detect the faults separately and then pass them to the LLM for troubleshooting steps. The LLM also better understands the problem using a specialized RAG created from the Packet Core Controller documents. The results were also promising, showing the system performs well when enough data is provided for each edge case. The proposed method will help enable more efficient testing and network management for high-end telecommunications systems.

Future works include adding more input forms to the model and not restricting it to PCAP files such as log dumps and network configuration files. These files will contain more detailed versions of the errors that can be used as context to the LLM to enable it to provide a better resolution. Other places that can be investigated includes using multimodal LLMs that will utilise the log files or PCAP files directly to the model without the need of a parser. This might enable faster inference times and more efficient training pipelines.